**Altermagnetic Signatures and Topological Textures in Ultrathin Multiferroic $BiFeO_3$**

Geo Fratian[1,27], Maya Ramesh[2,27], Xinyan Li[3,4,27], Evangelos Golias[5], Yousra Nahas[6], Sebastian Maria Ulrich Schultheis[7], Julian Skolaut[7], Marti Checa[8], Arundhati Ghosal[1], Evan Krysko[2,9], Paul Malinowski[10], Brendan D. Faeth[9,10], Jan Priessnitz[11], F. C. Fobasso Mbognou[11], Atsushi Hariki[12], Kosuke Sakurai[12], Mizuki Furo[12], Shashank Kumar Ojha[3,4], Morgan Congdon[1], Shiyu Zhou[13], Alexander Qualls[1], Kai Litzius[14], Christoph Klewe[15], Peter Meisenheimer[16], Jan Kunes[17], Kyle Shen [9,18], Laurent Bellaiche[6,19], Libor Šmejkal[11,20,21], Darrell G. Schlom[2,22,23], Yimo Han[3,4,24], Sergei Prokhorenko[6], Ramamoorthy Ramesh[16,25], Paul Stevenson[26], Angela Wittmann[7,*], Lucas Caretta[1,13*]

*[1]School of Engineering, Brown University, Providence, Rhode Island, USA*
*[2]Department of Materials Science and Engineering, Cornell University, Ithaca, New York, USA.*
*[3]Department of Materials Science and Nanoengineering, Rice University, Houston, TX 77005 USA*
*[4]Rice Advanced Materials Institute, Rice University, Houston, TX 77005 USA*
*[5]MAX IV Laboratory, Lund, Sweden*
*[6]Smart Ferroic Materials Center, Physics Department, Institute for Nano-science and Engineering, University of Arkansas, Fayetteville, Arkansas 72701, USA*
*[7]Institute of Physics, Johannes Gutenberg-University Mainz, Mainz, Germany*
*[8]Center for Nanophase Materials Sciences, Oak Ridge National Laboratory, Oak Ridge, TN, 37831, USA.*
*[9] Platform for the Accelerated Realization, Analysis, and Discovery of Interface Materials (PARADIM), Cornell University, Ithaca, NY, 14853, USA*
*[10] Department of Physics, Laboratory of Atomic and Solid State Physics, Cornell University, Ithaca, NY, 14853, USA*
*[11]Max Planck Institute for the Physics of Complex Systems, Nöthnitzer Str. 38, 01187 Dresden, Germany*
*[12]Department of Physics and Electronics, Graduate School of Engineering, Osaka Metropolitan University, 1-1 Gakuen-cho, Nakaku, Sakai, Osaka 599-8531, Japan*
*[13]Department of Physics, Brown University, Providence, Rhode Island, USA*
*[14]Experimental Physics V, Center for Electronic Correlations and Magnetism, University of Augsburg, Augsburg, Germany*
*[15]Advanced Light Source, Lawrence Berkeley National Laboratory, Berkeley, CA, 94720, USA.*
*[16]Department of Materials Science and Engineering, University of California, Berkeley, California, 94720, USA*
*[17]Department of Condensed Matter Physics, Faculty of Science, Masaryk University, Kotlářská 2, 61137 Brno, Czech Republic*
*[18]Kavli Institute at Cornell for Nanoscale Science, Cornell University, Ithaca, NY, 14853, USA*
*[19]Department of Materials Science and Engineering, Tel Aviv University, Ramat Aviv, Tel Aviv 6997801, Israel*
*[20]Max Planck Institute for Chemical Physics of Solids, Nöthnitzer Str. 40, 01187 Dresden, Germany*
*[21]Institute of Physics, Czech Academy of Sciences, Cukrovarnická 10, 162 00 Praha 6, Czech Republic*
*[22]Kavli Institute at Cornell for Nanoscale Science, Cornell University, Ithaca, NY, USA*
*[23]Leibniz-Institut für Kristallzüchtung, Berlin, Germany*
*[24]Smalley-Curl Institute, Rice University, Houston, TX 77005, USA*
*[25]Department of Physics, University of California, Berkeley, California, 94720, USA*
*[26]Department of Physics, Northeastern University, Boston, MA 02115 USA*
*[27]These authors contributed equally.*

*Correspondence to: lucas_caretta@brown.edu, a.wittmann@uni-mainz.de

**Abstract**

**Magnetoelectric multiferroics promise low-power switching, nonvolatile memory, and signal transduction, yet practical deployment requires ultrathin films where polar and magnetic order often degrade. Uncompensated bound charge drives depolarization in ferroelectrics, while off-stoichiometry, reduced anisotropy, and charge transfer can produce magnetic dead layers. Here, we realize a multiferroic phase of $BiFeO_3$ with no dead layer, revealing signatures of emergent altermagnetism in the ultrathin limit. First-principles calculations, symmetry analysis, atomic-resolution imaging, and magnetic imaging show that short-circuit electrostatic boundary conditions and epitaxial strain drive a continuous, thickness-induced phase transition accompanied by multiferroic topological textures. At four unit cells, these boundary conditions stabilize a collinear monoclinic phase with d-wave altermagnetic order, supported by magnetic circular dichroism and spin-resolved photoemission measurements. These results establish a pathway to stabilize unconventional multiferroicity at device-relevant thicknesses, reframing scaling limits for oxide electronics.**

## Main Text

## Introduction

Correlated thin film materials often suffer from loss of functionality at the thinnest length scales due to interface/surface effects, dimensionality, chemical intermixing, or even strain. This challenge is particularly pervasive in ferroic complex oxide perovskites. For example, magnetic dead layers are well-known in manganites[1–3] due to structural distortions and changes in interfacial electronic states. In thin ferroelectric films, large depolarizing fields from uncompensated bound charge suppress polarization, such as in $Pb(Zr_xTi_{1-x})O_3$ [4–7]. Multiferroic $BiFeO_3$ (BFO) is no exception to unfavorable thickness scaling, where both the antiferromagnetism and switchable polarization are reported to vanish below approximately ten unit cells[8,9]. This occurs primarily due to strong depolarizing electric fields resulting from BFO's large out-of-plane ferroelectric polarization and loss of magnetic anisotropy. These challenges in scaling have hindered the use of BFO in microelectronic technologies, as thinner multiferroic materials are crucial for low voltage devices. Beyond limiting device scaling, dead layers also restrict access to magnetic states that may otherwise emerge at reduced dimensions, including symmetry-driven unconventional phases such as altermagnetism. Several strategies have been pursued to retain polar order in the ultrathin regime of BFO and other ferroelectric oxides, including electrostatic techniques[6,10–13], chemical termination at ferroelectric interfaces[14,15], and confinement in superlattice structures[16]. Furthermore, to prevent magnetic dead layer formation, proximity/geometric effects[17,18] and strain[19] have been proposed as engineering strategies.

Despite these efforts, however, maintaining multiferroic functionality and liberating new phenomena in BFO at the thinnest length scales remains a fundamental challenge, as cooperative design of ferroelectric and magnetic order in multiferroics is challenging.

Here, we apply an atomic-scale design strategy that utilizes short-circuit boundary conditions and compressive strain induced by substrate-film lattice mismatch to stabilize ferroelectric and magnetic order in BFO with no dead layer. Furthermore, by preserving both order parameters in the ultrathin limit, this approach enables us to probe whether new coupled polar-magnetic phases, including altermagnetic states, can emerge once conventional thickness-scaling limitations are removed. We find that at thicknesses less than ~3 nm, BFO undergoes a continuous structural and ferroic phase transition which favors preferential out-of-plane polarization and coincides with the emergence of topological multiferroic textures. This rarely reported[20], emergent phase of BFO exhibits properties consistent with d-wave altermagnetism, as suggested by first-principle calculations, angle-resolved X-ray dichroism, and spin- and angle-resolved photoemission spectroscopy (spin-ARPES). Thus, by eliminating dead layers, we uncover a metastable ultrathin phase of BFO, which retains polar order and magnetic order, while also uncovering new topological phases and altermagnetism.

**Thickness induced phase transformation and retained polar order**

To retain polar and magnetic order in the ultrathin regime, we employ two design strategies. First, to suppress depolarizing fields, we synthesize BFO on a lattice matched oxide electrode whose free electrons provide sufficient electronic screening. Second, we impose large epitaxial compressive strain via a lattice mismatched substrate to stabilize out-of-plane structural distortions and, consequently, out-of-plane polarization. We realize both by synthesizing BFO thin films on (110)-oriented $DyScO_3$ (DSO) substrates with 20 nm thick $SrRuO_3$ (SRO) bottom electrodes using reactive oxide molecular-beam epitaxy (Methods). Although the BFO/SRO//DSO heterostructure has been widely studied in various contexts, the mismatch in octahedral tilt patterns between BFO and SRO can induce interfacial structural frustration that becomes dominant at sufficiently small thicknesses, with implications for the thickness scaling of ferroic order[21–23].

High-angle annular dark-field scanning transmission electron microscopy (HAADF-STEM ) images of a heterostructure containing a four-unit-cell-thick BFO layer (Fig. 1a,b), acquired along the $[1\bar{1}0]_O$ and $[001]_O$ zone axes of the DSO substrate (O denotes orthorhombic), demonstrate single crystal films with atomically sharp interfaces. Single phase films and atomically smooth surfaces are confirmed with X-ray diffraction and atomic force microscopy (Extended Data Fig. 1, Supplementary Fig. 1). Multislice electron ptychography was performed to obtain super-resolution phase images for polar mapping in the ultrathin BFO layer (Fig. 1a,b, Methods). Remarkably, relative to bulk or thicker BFO films, where BFO adopts a rhombohedral *R3c* structure with spontaneous ferroelectric polarization **P** aligned along $<111>_{PC}$ (PC denotes pseudocubic), the ultrathin films exhibit a modified symmetry with an *enhanced* out-of-plane-to-in-plane polarization ratio, precisely where depolarizing fields are typically largest, yielding a *Cm* symmetry. This effect is most pronounced along the more compressed $[010]_{PC}$ film zone axis, where the film exhibits a purely out-of-plane $[001]_{PC}$ polarization (Fig. 1c,d). Notably, we observe no ferroelectric dead layer in the BFO, implying that the first deposited unit cell under these boundary conditions is polar. Moreover, along both zone axes, BFO also induces symmetry breaking (cation displacement) in the adjacent SRO underlayer consistent with a polar metal, which may in turn help stabilize the ferroelectric state in the first unit cell of BFO[24] and help avoid significant polar discontinuities previously reported at similar interfaces[25–27]. Additional STEM imaging of other BFO thicknesses is shown in Supplementary Figs. 2 and 3, along with analyses of the SRO/BFO interface structure (Supplementary Discussion I). Using quadrature phase differential interferometer (QPDI)-based piezoresponse force microscopy (PFM), we find indications of bistable polar order (hysteresis), suggesting the ability to switch the polar order is preserved. (Extended Data Fig. 2).

To better understand the origin of the observed monoclinic structure, we performed first-principles calculations for the BFO/SRO bilayer geometry (Methods; Extended Data Fig. 3, Supplementary Discussion I). Strained BFO/SRO bilayer calculations reproduce the key experimental features of the ultrathin films, including the two-component polarization, suppressed octahedral rotation along one in-plane direction, and enhanced out-of-plane-to-in-plane polar displacement near the top BFO interface.

These results indicate that the SRO bottom electrode provides sufficient screening to retain out-of-plane polarization in atomically thin BFO, while interfacial structural constraints further select the observed monoclinic symmetry. We therefore attribute the stabilization of the *Cm* phase to the combined action of anisotropic compressive strain from DSO, electrostatic screening from SRO, constrained octahedral tilts at the BFO/SRO interface, and symmetry breaking at the BFO/vacuum interface, which together drive BFO along the $R$-$M_a$-$M_c$-$T$ pathway without fully reaching the tetragonal phase.

**Magnetic order in ultrathin BFO**

Having established boundary conditions that stabilize polar order down to the first unit cell, we next examine how these conditions, and the associated structural phase transition, affect the magnetic properties in ultrathin *Cm* BFO. In BFO, the ferroelectric polarization induces a strong Dzyaloshinskii–Moriya interaction (DMI) that typically results in a incommensurate spin cycloid oriented along $\langle 1\bar{1}0\rangle_{\mathrm{PC}}$ or $\langle 11\bar{2}\rangle_{\mathrm{PC}}$ perpendicular to **P**[28–30]. The cycloid propagation vector can be tuned via epitaxial strain, electrostatic boundary conditions, and applied electric fields[31,32]. As our engineered boundary conditions stabilize a unique phase of BFO, it could be expected that the magnetic properties are significantly different than in bulk. To probe these effects, we employ X-ray linear dichroism (XLD) in total electron yield mode to examine global changes in magnetic order and scanning nitrogen-vacancy center (NV) magnetometry to image local changes. First, we use XLD with differential polarization contrast (Methods). In BFO, linear dichroism signal can be sensitive to both AFM and polar order depending on the chosen elemental resonance[33–35]. Specifically, we measure at the Fe $L_2$-edge, where linear dichroism in BFO is predominantly sensitive to the AFM order[33,36–39] (Supplementary Discussion II, Extended Data Figs. 4-5, Supplementary Fig. 5). Figures 2a-c present the Fe $L_2$-edge XLD intensity as a function of BFO thickness in BFO/SRO//DSO heterostructures for several crystallographic directions (see insets; Extended Data Fig. 6). First, we measure under grazing-incidence illumination (16° relative to the sample plane), with X-rays incident along $[010]_{\mathrm{PC}}$ film direction and *s*- and *p*-polarized light oriented parallel and perpendicular to the sample plane, respectively (Fig. 2a inset). Here, Fe $L_2$-edge XLD signal is reduced for thicknesses below

~10 nm, consistent with prior reports on BFO at these scales[8]. However, by contrast, the XLD intensity along $[100]_{PC}$ significantly *increases* for BFO films thinner than ~10 nm (Fig. 2b). These trends indicate that AFM order persists at the thinnest length scale with stronger magnetic anisotropy towards $[010]_{PC}$.

The STEM imaging (Fig. 1) shows that the polarization is largely out of plane in the ultrathin regime. Normal-incidence XLD is not sensitive to out-of-plane polar order. Thus, XLD should have little sensitivity to ferroelectric order, allowing us to exclude a significant ferroelectric contribution to the measured signal. Normal-incidence XLD data is shown in Fig. 2c, confirming this pronounced in-plane AFM anisotropy along $[010]_{PC}$ with a Néel temperature near 500 K in 1.6 nm (4 u.c.) thick films and 350 K in 0.4 nm (1 u.c.) thick films (Extended Data Fig. 6). The continuous, thickness-dependent rotation of the magnetic easy axis is consistent with a second-order phase transformation that coincides with the structural transition observed by STEM.

To investigate the evolution of magnetic order at ultrathin thickness, we use spatially-resolved scanning NV magnetometry[40] (Methods; Fig. 2d) to map local stray fields from the sample. In films ≥10 nm thickness, NV mapping shows the characteristic zigzag pattern of a Type-I spin cycloid in BFO, consistent with prior reports[31,32,40]. At 5 nm, near the thickness-driven structural transition, vestiges of a periodic cycloidal texture remain, but the propagation length is markedly reduced, and a clear zigzag pattern is absent. At 1.6 nm (four unit cells), the NV signal lacks typical signatures of cycloidal order and instead more resembles weak ferromagnetism, which we confirm later with XLD imaging and DFT. By examining the persistence of the long-range order via a Fast Fourier Transform across the thickness series (Supplementary Figs. 6-7), we confirm that magnetic order persists to the ultrathin limit with indications of cycloid suppression below 5 nm.

**Topological multiferroic defects across a continuous phase transition**

To track the evolution of the ferroelectric and magnetic domain structure through the thickness-driven phase transformation, we map domain structures using angle-resolved XLD photoemission electron microscopy (PEEM; Methods). Contrasting with the XLD above, XLD-PEEM images were acquired at the Fe $L_3$-edge, which, in addition to providing higher signal in BFO, gives nearly equal contrast from polar and magnetic order[33–35] (Supplementary Discussion II; Extended Data Figs. 4-5, Supplementary Fig. 5). Images were acquired under normal-incidence X-ray illumination, which probes in-plane projections of polar and AFM components. Figures 3a-c show XLD PEEM vector maps (Methods) at representative BFO thicknesses – 50 nm, 3.5 nm, and 1.6 nm – across the structural phase transition. In 50 nm thick films under these boundary conditions, BFO is expected to relax into a striped ferroelectric domain pattern separated by 71° ferroelectric domain walls[41]. Consistent with this, Fig. 3a displays alternating red/blue contrast from the in-plane component of the striped polar order, and piezoresponse force microscopy images (Supplementary Fig. 8), as well as Oxygen K-edge XLD[34,35] (Supplementary Discussion II; Extended Data Figs. 4-5, Supplementary Fig. 5) corroborate that these stripes correspond to ferroelectric domains. We note that 180° domains exhibit identical contrast levels and are indistinguishable via XLD alone. Despite confirming its appearance in NV magnetometry imaging (Fig. 2d), the spin cycloid is not resolved at this thickness, indicating that the XLD contrast in the 50 nm is dominated by the in-plane component of ferroelectric order in the thick-film limit. Radial histograms in Fig. 3d and Extended Data Fig. 4 quantify the angular distribution of order parameter orientation, where red/blue indicates ferroelectric order, while purple/white indicates AFM order. The histograms confirm ferroelectric-dominated signal in thick BFO (Supplementary Discussion III). Nevertheless, pronounced magnetic contrast appears locally near ferroelastic domain boundaries, consistent with emergent magnetic behavior at BFO domain walls reported elsewhere[42–44].

At reduced thickness, BFO undergoes pronounced domain reconfiguration, as seen in the 3.5-nm XLD-PEEM map (Fig. 3b). The normal-incidence imaging geometry only probes in-plane components; consequently, as the polarization rotates toward the out-of-plane direction $[001]_{PC}$ (Figs. 1a,b), AFM moments rotate in-plane, as they are fixed normal to **P**, resulting in a reduction of the ferroelectric contrast and an underlying magnetic texture that emerges (Supplementary Fig. 7). The ferroelectric pattern (red/blue) loses its long-range stripe character and becomes mosaic-like. Short-range purple/white stripes within each polar domain trace the in-plane component of the spin-cycloid Néel vector, with a periodicity of ~71 nm (Supplementary Fig. 9), consistent with prior reports for BFO[31,32,40,45]. The ~90° rotation of the projected spin cycloid between neighboring ferroelectric domains is also consistent with the Type-I cycloid expected in BFO[31,32,40], as the cycloid propagation vector $\boldsymbol{k}$ is locked to the ferroelectric domain orientation (i.e., for a given $\mathbf{P}$, $\boldsymbol{k}$ remains fixed), indicating retained magnetoelectric coupling at this thickness. Notably, here we resolve the spin cycloid in BFO using standard XLD PEEM imaging, circumventing more complicated techniques and sample preparation, such as soft X-ray ptychography[35]. Importantly, at the boundaries between mosaic ferroelectric domains, polar and magnetic order become frustrated, producing a rich landscape of topological multiferroic textures (pink and green boxes in Fig. 3b). Enlarged vector maps of representative textures are shown in Fig. 3c–e with accompanying schematics. The local vorticity, defined as $\boldsymbol{\omega} = \nabla \times \boldsymbol{u}$ with $\boldsymbol{u}$ the order-parameter field (polar or Néel), is finite in these regions, yielding vortex-like and vortex-ring configurations comprising of both polarization and AFM order, whose finite winding corresponds to a nonzero topological charge. The topological features appear at ~10 nm and their density increases steadily as thickness is reduced to 3.5 nm. (Extended Data Fig. 7). The emergence of these thickness-driven textures is consistent with Kibble-Zurek physics, where a continuous transition "freeze in" domains of polar and Néel order that later unwind into vortex configurations[46,47], also seen in other AFMs[48]. Notably, within these frustrated multiferroic regions, $\mathbf{P}$ and $\boldsymbol{k}$ are not directly correlated; instead, they appear to jointly constitute the topological textures.

At 4 unit cells thickness, the phase transformation is complete, rendering a distinct XLD vector map. At this thickness, $\mathbf{P}$ is predominately out-of-plane, yielding diminished in-plane polar contrast and signal that is dominated by the in-plane component of the magnetization (purple/white in Fig. 3f). However, unlike thicker films, this phase of BFO does not contain cycloidal order, but rather appears to host canted *collinear* antiferromagnetic order that is highly anisotropic, consistent with NV magnetometry (Fig. 2d). As expected from macroscopic XLD measurements (Fig. 2), AFM domains predominately have Néel vector orientations along $[010]_{PC}$ (purple) with smaller domains along $[100]_{PC}$ (white). This domain pattern is similar to that measured in hematite ($\alpha$-$Fe_2O_3$), the parent compound of BFO, near its Morin transition[48,49].

**Emergent altermagnetic signatures in ultrathin BFO**

By establishing boundary conditions that eliminate the dead layer in BFO, we reveal a previously hidden non-cycloidal magnetic state with *Cm* symmetry. In conventional bulk BFO, nonrelativistic spin splitting has been predicted for several stable phases, provided that collinear magnetic order is present, rather than cycloidal order commonly found in BFO[50–53]. The suppression of the spin cycloid and reduction of crystal symmetry in our ultrathin phase of BFO therefore suggests emergent altermagnetic behavior. At the collinear spins level, our DFT calculations predict a G-type AFM ground state ordering of Fe spins for the *Cm* BFO phase shown in Fig. 4a together with calculations of the altermagnetic spin density, which is consistent with our spin group theory analysis (Methods). The spin group generators include a combined symmetry of spin rotation and mirror plane[54] along the $[001]_{PC}$-crystallographic direction, $[C_2||m_b]$ marked in Fig. 4a, and lattice centering operation $[E||E|(½\ ½\ 0)]$. The corresponding spin Laue group is $^{2}2/^{2}m$ resulting in a bulk d-wave altermagnetism of B-2 class[54]. The B-2 class d-wave altermagnetism in our *Cm* BFO is characterized by an altermagnetic nodal plane protected by the $[C_2||m_b]$, and another nodal plane that is not enforced to be flat due to the low symmetry of the *Cm* BFO phase. This is distinct from the previously studied bulk BFO with predicted g-wave and i-wave altermagnetic spin splitting[50,51].

In Figs. 4b,c, we study the corresponding altermagnetic spin polarization and spin splitting in momentum space. The nonrelativistic (spin orbit coupling switched off) DFT calculations confirm the d-wave altermagnetic symmetry of spin polarization in *Cm* BFO as seen on the constant energy isosurface in Fig. 4b. Figure 4c shows the corresponding alternating nonrelativistic spin splitting of the bands along the *-Y'ΓY'* and *-YΓY* directions. The magnitude of the spin splitting is most noticeable in the vicinity of the valence band maximum reaching up to ~0.3 eV. This nonrelativistic spin splitting magnitude is largest among the hitherto predicted altermagnetic multiferroic candidates[51,55]. To experimentally probe the altermagnetic spin splitting of the band structure in the metastable *Cm* phase, we performed spin-ARPES (Methods, Extended Data Fig. 8). Although BFO is generally insulating, the ultrathin nature of the film, combined with the thick conductive SRO bottom electrode, helps mitigate some of the detrimental effects of charging during measurement. Moreover, to preserve surface quality, the samples were transferred *in-* *vacuo* to the spin-ARPES system immediately following oxide-MBE deposition (Methods). Fig. 4d presents the corresponding spin-resolved iso-energy map for the momentum slice indicated in Fig. 4b. The spin-resolved iso-energy map reveals a momentum-dependent reversal of $P_x$, consistent with the calculated altermagnetic spin texture on the corresponding constant-energy surface. Although the present data do not directly resolve separate spin-dependent band branches, the observed spin-polarization reversal provides supporting evidence for altermagnetic order in the *Cm* phase.

X-ray spectroscopy circular dichroism can further provide local experimental insights into altermagnetic behavior[56–58] (Supplementary Discussion IV). Extended Data Fig. 9 shows DFT-calculated Fe L-edge absorption and X-ray magnetic circular dichroism (XMCD) spectra for the monoclinic *Cm* phase of BFO with several Néel vector orientations and incident beam directions (Methods). The simulations reveal a strong XMCD response arising from non-relativistic altermagnetism, with a spectral line shape that is qualitatively distinct from that of any spin-orbit-driven effects in BFO (e.g., the weak canted moment), which contributes negligibly to the XMCD signal (Supplementary Fig. 10). We leverage this distinct altermagnetic spectral fingerprint to aid in distinguishing altermagnetic and canted moment XMCD near the Fe $L_3$-edge (Supplemental Discussion V, Extended Data Fig. 10, Supplementary Fig. 11). Figures 4e and f plot co-registered local XLD vector and XMCD images from the same region of the 1.6 nm film. Vector-XLD maps in Fig. 4e show the real space orientation of the compensated Néel vector configuration, while XMCD records the time-reversal symmetry breaking permitted by altermagnetic symmetry (Supplementary Discussion V-VII). In Fig. 4f, pronounced XMCD contrast emerges in the thinnest, collinear, monoclinic sample, whereas its presence is absent at all other thicknesses >1.6 nm when the spin cycloid is still present (Extended Data Fig. 10, Supplementary Fig. 12, Supplementary Discussion VI). We note that in addition to its negligible calculated contribution to the signal (Extended Data Fig. 10), monoclinic BFO's weak canted moment is predicted to be largely in-plane and thus falls outside the sensitivity of our out-of-plane XMCD geometry (Supplementary Discussion VII). Combining the two channels through a logical (Boolean) reconstruction yields a full real-space map of the Néel vector[57,58], a capability that arises from the interplay between altermagnetic symmetry and Hall conductivity, and links real-space imaging to reciprocal-space band structure. Analogous to the Hall effect in transport measurements, XMCD can be cast in terms of a Hall pseudovector[59] (Supplemental Discussion VI). In our system, the $[100]_{PC}$ orientation (perpendicular to the mirror plane) of the Néel vector combined with the observed BFO structure (Fig. 1) is compatible with the *Cm'* magnetic symmetry or *m*' magnetic point group. The latter allows for the Hall pseudovector lying within the mirror plane (Fig. 4a) containing ferroelectric polarization. While Fig. 4e plots the detection of the axis of the Néel vector, the absolute direction remains unresolved. This information is revealed by combining the XMCD- and XMLD-PEEM in a colored vector map shown in Fig. 4g, which plots this Boolean combination from Figs. 4e and f. The result reveals the underlying altermagnetic Hall pseudovector in the real-space domain structure of the ultrathin BFO (Supplemental Discussion VIII). As anticipated, the 4-fold symmetry of the domains in Fig. 4e is consistent with the d-wave altermagnetic symmetry from DFT (Figs. 4a and b). We note that in the normal-incidence geometry used here, symmetry constraints prevent distinguishing altermagnetic vectors when the Néel vector lies within the mirror plane (Supplementary Discussion VII, Extended Data Figs. 9-10, Supplementary Figs. 10-11), as indicated by the purple contrast in Fig. 4g.

**Outlook**

We present a design and synthesis pathway that preserves room-temperature multiferroic order down to ultrathin thicknesses. Polar-vector mapping reveals an enhanced out-of-plane to in-plane polarization ratio at the smallest scales. We observe a continuous, thickness-driven phase transition consistent with Kibble-Zurek-like behavior, during which multiferroic topological textures emerge. At four unit cells, a collinear monoclinic *Cm* state is stabilized in which the cycloid is quenched and strong evidence of d-wave altermagnetism appears[51]. Taken together, these measurements provide a picture of real-space multiferroic topology at device-relevant thicknesses with d-wave altermagnetic signatures in the XMCD, DFT, spin-ARPES, and symmetry analysis. Key next steps include directly demonstrating voltage-driven reversal of the altermagnetic band splitting and multiferroic topologies via ferroelectric switching. While initial attempts are presented in Supplementary Discussion III and Supplementary Fig. 13-14, rapid de-polling in the present heterostructure limits a conclusive demonstration, motivating future efforts in more symmetric device architectures. Additional detector-magnetization reversal and spin-ARPES energy distribution curves would further confirm the altermagnetic spin splitting. Beyond this specific material, the approach outlines a general blueprint for engineering ultrathin multiferroicity with multi-order-parameter topological textures, enabling voltage-based writing with symmetry-selective readout.

**Acknowledgements:**

Devices and films were processed using equipment in the Brown University Instrumentation for Molecular and Nanoscale Innovation facility. This research used resources of the Advanced Light Source, a U.S. DOE Office of Science User Facility under contract no. DE-AC02-05CH11231. This work made use of the synthesis facility of the Platform for the Accelerated Realization, Analysis, and Discovery of Interface Materials (PARADIM), which is supported by the National Science Foundation under Cooperative Agreement No. DMR-2039380. The authors thank T. Butcher and M. Furo for valuable discussion.

**Funding Statement**:

Scanning nitrogen vacancy magnetometry measurements were supported by the Air Force Office of Scientific Research under FA9550-24-1-0169. $BiFeO_3$ synthesis and measurements were supported by NSF EPSCOR RII Track-4 Research Fellows Program under OIA-2327352 and under ETHOS MURI via cooperative agreement W911NF-21-2-0162. X.L. acknowledges support from the Rice Advanced Materials Institute (RAMI) at Rice University as a RAMI Postdoctoral Fellow. X.L. and Y.H. acknowledge support from NSF (FUSE-2329111 and CMMI-2239545) and Welch Foundation (C-2065). X.L. and Y.H. acknowledge the Electron Microscopy Center, Rice University. S. M. U. S., J. S., and A.W. acknowledge funding from the Deutsche Forschungsgemeinschaft (DFG, German Research Foundation) projects 268565370 (SFB TRR173 projects A09 and B14) and 465145163 (CRC 1552 project A02). LŠ acknowledges funding from the ERC Starting Grant No. 101165122 and Deutsche Forschungs-gemeinschaft (DFG) grant no. TRR 288 – 7422213477 (Projects A09 and B05). Quadrature phase differential interferometer-based PFM measurements were conducted as part of a user project at the Center for Nanophase Materials Sciences (CNMS), which is a US Department of Energy, Office of Science User Facility at Oak Ridge National Laboratory. L.B. also acknowledges the support of the Vannevar Bush Faculty Fellowship (VBFF) Grant No. N00014-20-1-2834 from the Department of Defense and ARA Impact Grant 3.0. We thank MAX IV Laboratory for time on beamline MaxPEEM under proposals 20230367, 20240599, 20241342, 20250678, 20250994 and 20251711. Research conducted at MAX IV, a

Swedish national user facility, is supported by the Swedish Research Council under contract 2018-07152, the Swedish Governmental Agency for Innovation Systems under contract 2018-04969, and Formas under contract 2019-02496. M.C. acknowledges that this material is based upon work supported by the National Science Foundation Graduate Research Fellowship Program under Grant No.2439559. Any opinions, findings, and conclusions or recommendations expressed in this material are those of the author(s) and do not necessarily reflect the views of the National Science Foundation. This work was also supported by JSPS KAKENHI Grant Numbers 25K00961, 25K07211, (A.H.), the project Quantum materials for applications in sustainable technologies (QM4ST), funded as project No.~CZ.02.01.01/00/22 008/0004572 by Programme Johannes Amos Commenius, call Excellent Research (J.K.) and by the Ministry of Education, Youth and Sports of the Czech Republic through the e-INFRA CZ (ID:90254) (A.H., J.K.). Part of the computations in this work were performed using the facilities of the Supercomputer Center, the Institute for Solid State Physics, the University of Tokyo.

**Author Contributions Statement:** L.C. led study and guided the team.; G.F., E.G., J.S., K.L, C.K., M.C., A.W, and L.C. performed all X-ray spectroscopy and imaging measurements.; X-ray spectroscopy was analyzed by G.F., E.G., S.M.U.S, A.W., under guidance from L.C. and A.W.; M.R. optimized the synthesis of BFO thin films and performed X-ray diffraction studies under guidance from D.G.S. and L.C.; X.L. performed all STEM measurements and analysis under the supervision of Y.H.; A.G. and M.R. performed NV magnetometry and PFM with the help of S.K.O., S.Z., and A.Q., under the supervision of R.R., P.S., and L.C. Quadrature phase differential interferometer-based PFM was performed by M.C.; First principles calculations were performed by S.P. and Y.N. under guidance from L.B., as well as J.P. and F.C.F.M. under supervision of L.S.; XMCD calculations were performed by A.H., K.S., M.F., and J.K.; M.R., P.M., E.K., B.D.F., performed and analyzed spin-ARPES measurements under the supervision of K.S.; The manuscript was prepared by L.C., A.W., G.F., E.G., P.M., P.S., M.R., X.L., L.S., S.P., L.B., and R.R.; All authors contributed to the discussion of the data in the manuscript and the supplementary materials.

**Competing interests Statement:** The authors declare no competing interests.

**Figure Captions (for Main Text Figures):**

**Figure 1 | Monoclinic structure and retained polar order in ultrathin $BiFeO_3$ (BFO).** Atomically resolved high-angle annular dark-field scanning transmission electron microscopy (HAADF-STEM) and multislice electron ptychography-based polar vector mapping in 4 unit cell BFO grown on $DyScO_3$ with $SrRuO_3$ (SRO) bottom electrodes along the **a)** $[010]_{PC}$ and **b)** $[100]_{PC}$ zone axes. Scale bars in (a) and (b) are 1 nm. **c,d)** Out-of-plane and in-plane atomic displacement mapping of the ptychographic phase images shown in (a) and (b), respectively, across the BFO/SRO interface. No ferroelectric dead layer and induced polarization in the SRO is observed. Each data point represents the mean displacement of atomic sites within the corresponding unit-cell row, and error bars indicate ±1 standard deviation. The numbers of atomic sites (n) for unit-cell sequences 1–7 are 8, 11, 12, 11, 10, 9, and 6 in (c), and 9, 9, 9, 8, 7, 5, and 3 in (d), respectively. **e,f)** Expanded view of the unit cell of BFO (SRO), illustrating the octahedral tilt pattern present (absent) along each in plane zone axis. Scale bars in (e) and (f) are 1 Å. **g)** Unit cell of the relaxed *Cm* phase obtained from DFT simulations with bulk BFO settings.

**Figure 2 | Retained collinear magnetic order in ultrathin $BiFeO_3$ (BFO).** Fe $L_2$ edge X-ray linear dichroism (XLD) at grazing incidence (16°) along the **a)** $[010]_{PC}$ and **b)** $[100]_{PC}$ sample directions as a function of thickness in BFO/$SrRuO_3$(20 nm)//DSO heterostructures. Fe $L_2$ edge XLD in BFO is predominately sensitive to magnetic order. **c)** Fe $L_2$ edge XLD thickness scaling at normal incidence in the same heterostructure as (a) and (b). Insets show the orientation of X-ray light polarization relative to the sample. **d)** High resolution scanning nitrogen vacancy magnetometry as a function of BFO thickness. The 1.6 nm sample was probed using pulsed optically detected magnetic resonance, whereas the rest correspond to iso-magnetic field imaging (Methods). The scale bars in (d) are 500 nm, except the 1.6 nm sample, where it is 250 nm. a.u., arbitrary units; kc/s, kilocounts per second.

**Figure 3 | Emergent multiferroic topology in thin $BiFeO_3$ (BFO). a)** Fe $L_3$ edge normal incidence X-ray linear dichroism – photo emission electron microscopy (XLD-PEEM) in-plane vector map of **a)** 50 nm, **b)** 3.5 nm, and **c)** 1.6 nm of $BiFeO_3$ grown on $DyScO_3$ with $SrRuO_3$ bottom electrodes. Fe $L_3$ edge XLD in BFO is sensitive to both polar and antiferromagnetic (AFM) order. Inset color wheels show the orientation of each order parameter with red/blue contrast corresponding to polar order and purple/white contrast corresponding to AFM order. Scale bars in (a)-(c) are 1 μm. **d)** Order parameter radial histograms extracted from the fit of the vector maps for various thicknesses (*t*) of BFO (Methods). **e-g)** Exemplary topological defects from colored boxes in (b) are highlighted, showing vertex and vertex rings forming during the continuous phase transition. The scale bars in (e)-(g) are 200 nm.

**Figure 4 | Altermagnetic signatures in ultrathin $BiFeO_3$ (BFO). a)** Crystallographic structure and **b)** nonrelativistic altermagnetic spin-splitting in the band structure calculated from DFT of *Cm* BFO. The spin splitting (opposite spin channels shown by magenta and cyan color) confirms the altermagnetic d-wave symmetry, as it alternates in sign along the *-Y'ΓY'* and *-YΓY* directions (notation marked in the Brillouin), and time-reversal symmetry breaking, as the spin is the same for opposite wavevectors. The altermagnetic $\mathcal{T}$ breaking allows for XMCD and Hall vector **h** in the mirror plane when the Néel vector **L** is oriented along $[100]_{PC}$ direction as marked in panel (a). $E_F$, Fermi Energy. **c)** DFT calculated constant energy

isosurfaces in Brillouin zone at -0.08 eV from the top of the valence band. The spin polarization calculated on top of the constant energy isosurface shows altermagnetic d-wave symmetry. The opposite spin channels are related by altermagnetic spin group symmetry $[C_2||M_b]$, combining two-fold spin rotation $C_2$ with a black mirror plane $M_b$ along the crystallographic *b*-direction. **d)** Spin-polarized iso-energy map for the slice (d) shown in panel (c) experimentally measured via *in-vacuo* spin-ARPES. **e)** XLD-PEEM vector map and **f)** XMCD PEEM imaging of the 1.6 nm BFO sample taken at normal incidence at the Fe $L_3$ edge (Methods, Supplemental Discussion V-VI). **g)** XMCD binary mask of the XMLD vector map with green and orange domains corresponding to absolute orientations of the Néel vector along $[100]_{PC}$, while purple domains along $[010]_{PC}$ remain indistinguishable due to symmetry constraints. All scale bars are 0.5 μm.

## References

1. Park, J. H. *et al.* Magnetic Properties at Surface Boundary of a Half-Metallic Ferromagnet La0.7Sr0.3MnO3. *Phys. Rev. Lett.* **81**, 1953 (1998).
2. Sun, J. Z., Abraham, D. W., Rao, R. A. & Eom, C. B. Thickness-dependent magnetotransport in ultrathin manganite films. *Appl. Phys. Lett.* **74**, 3017–3019 (1999).
3. Borges, R. P., Guichard, W., Lunney, J. G., Coey, J. M. D. & Ott, F. Magnetic and electric "dead" layers in (La0.7Sr0.3)MnO3 thin films. *J. Appl. Phys.* **89**, 3868–3873 (2001).
4. Mehta, R. R., Silverman, B. D. & Jacobs, J. T. Depolarization fields in thin ferroelectric films. *J. Appl. Phys.* **44**, 3379–3385 (1973).
5. Fong, D. D. *et al.* Ferroelectricity in ultrathin perovskite films. *Science* **304**, 1650–1653 (2004).
6. Junquera, J. & Ghosez, P. Critical thickness for ferroelectricity in perovskite ultrathin films. *Nat. 2003 4226931* **422**, 506–509 (2003).
7. Tybell, T., Ahn, C. H. & Triscone, J. M. Ferroelectricity in thin perovskite films. *Appl. Phys. Lett.* **75**, 856–858 (1999).
8. Maksymovych, P. *et al.* Ultrathin limit and dead-layer effects in local polarization switching of BiFeO3. *Phys. Rev. B* **85**, 014119 (2012).
9. Huijben, M. *et al.* Ultrathin Limit of Exchange Bias Coupling at Oxide Multiferroic/Ferromagnetic Interfaces. *Adv. Mater.* **25**, 4739–4745 (2013).
10. Gradauskaite, E. *et al.* Defeating depolarizing fields with artificial flux closure in ultrathin ferroelectrics. *Nat. Mater. 2023* 1–7 (2023) doi:10.1038/s41563-023-01674-2.
11. Stengel, M., Vanderbilt, D. & Spaldin, N. A. Enhancement of ferroelectricity at metal-oxide interfaces. *Nat. Mater.* **8**, 392–397 (2009).
12. Béa, H. *et al.* Ferroelectricity Down to at Least 2 nm in Multiferroic BiFeO3 Epitaxial Thin Films. *Jpn. J. Appl. Phys.* **45**, L187 (2006).
13. Ji, D. *et al.* Freestanding crystalline oxide perovskites down to the monolayer limit. *Nature* **570**, 87–90 (2019).

14. Strkalj, N. *et al.* In-situ monitoring of interface proximity effects in ultrathin ferroelectrics. *Nat. Commun. 2020 111* **11**, 1–6 (2020).

15. Wang, H. *et al.* Direct observation of room-temperature out-of-plane ferroelectricity and tunneling electroresistance at the two-dimensional limit. *Nat. Commun. 2018 91* **9**, 3319- (2018).

16. Mundy, J. A. *et al.* Atomically engineered ferroic layers yield a room-temperature magnetoelectric multiferroic. *Nature* **537**, 523–527 (2016).

17. Ye, B. *et al.* Reducing the magnetic dead layer to one unit cell in ultrathin films of manganites using spin-orbit coupling. *Phys. Rev. B* **108**, 094406 (2023).

18. Guo, E. J. *et al.* Removal of the Magnetic Dead Layer by Geometric Design. *Adv. Funct. Mater.* **28**, 1800922 (2018).

19. Chen, Y. *et al.* Significant Reduction of the Dead Layers by the Strain Release in La0.7Sr0.3MnO3 Heterostructures. *ACS Appl. Mater. Interfaces* **14**, 39673–39678 (2022).

20. Dupé, B. *et al.* Competing phases in BiFeO3 thin films under compressive epitaxial strain. *Phys. Rev. B* **81**, 144128 (2010).

21. Heron, J. T. *et al.* Deterministic switching of ferromagnetism at room temperature using an electric field. *Nature* **516**, 370–373 (2014).

22. Rondinelli, J. M., May, S. J. & Freeland, J. W. Control of octahedral connectivity in perovskite oxide heterostructures: An emerging route to multifunctional materials discovery. *MRS Bull.* **37**, 261–270 (2012).

23. Lee, S. S. *et al.* Correlation between Geometrically Induced Oxygen Octahedral Tilts and Multiferroic Behaviors in BiFeO3 Films. *Adv. Funct. Mater.* **28**, 1800839 (2018).

24. Puggioni, D., Giovannetti, G. & Rondinelli, J. M. Polar metals as electrodes to suppress the critical-thickness limit in ferroelectric nanocapacitors. *J. Appl. Phys.* **124**, 55 (2018).

25. Nakagawa, N., Hwang, H. Y. & Muller, D. A. Why some interfaces cannot be sharp. *Nat. Mater. 2006 53* **5**, 204–209 (2006).

26. Ohtomo, A. & Hwang, H. Y. A high-mobility electron gas at the LaAlO3/SrTiO3 heterointerface.

*Nature* **427**, 423–426 (2004).

27. Yu, P. *et al.* Interface control of bulk ferroelectric polarization. *Proc. Natl. Acad. Sci. U. S. A.* **109**, 9710–9715 (2012).

28. Rahmedov, D., Wang, D., Añiguez, J. & Bellaiche, L. Magnetic cycloid of BiFeO3 from atomistic simulations. *Phys. Rev. Lett.* **109**, 037207 (2012).

29. Ramazanoglu, M. *et al.* Local weak ferromagnetism in single-crystalline ferroelectric BiFeO3. *Phys. Rev. Lett.* **107**, 207206 (2011).

30. Kadomtseva, A. M., Zvezdin, A. K., Popov, Y. F., Pyatakov, A. P. & Vorob'Ev, G. P. Space-time parity violation and magnetoelectric interactions in antiferromagnets. *JETP Lett.* **79**, 571–581 (2004).

31. Haykal, A. *et al.* Antiferromagnetic textures in BiFeO3 controlled by strain and electric field. *Nat. Commun.* **11**, 1704 (2020).

32. Meisenheimer, P. *et al.* Switching the spin cycloid in BiFeO3 with an electric field. *Nat. Commun.* **15**, 2903 (2024).

33. Zhao, T. *et al.* Electrical control of antiferromagnetic domains in multiferroic BiFeO 3 films at room temperature. *Nat. Mater.* **5**, 823–829 (2006).

34. Moubah, R. *et al.* Direct imaging of both ferroelectric and antiferromagnetic domains in multiferroic BiFeO3 single crystal using x-ray photoemission electron microscopy. *Appl. Phys. Lett.* **100**, 15 (2012).

35. Butcher, T. A. *et al.* Ptychographic Nanoscale Imaging of the Magnetoelectric Coupling in Freestanding BiFeO3. *Adv. Mater.* **36**, 2311157 (2024).

36. Kuo, C. Y. *et al.* Single-domain multiferroic BiFeO3 films. *Nat. Commun.* **7**, 12712 (2016).

37. Chen, Z. *et al.* Complex strain evolution of polar and magnetic order in multiferroic BiFeO3 thin films. *Nat. Commun.* **9**, 3764 (2018).

38. Yang, J. C. *et al.* Electrically enhanced magnetization in highly strained BiFeO3 films. *NPG Asia Mater. 2016 85* **8**, e269–e269 (2016).

39. Ko, K. T. *et al.* Concurrent transition of ferroelectric and magnetic ordering near room temperature. *Nat. Commun.* **2**, 567 (2011).

40. Gross, I. *et al.* Real-space imaging of non-collinear antiferromagnetic order with a single-spin magnetometer. *Nature* **549**, 252–256 (2017).

41. Ying-Hao Chu, B. *et al.* Nanoscale Domain Control in Multiferroic BiFeO3 Thin Films. *Adv. Mater.* **18**, 2307–2311 (2006).

42. He, Q. *et al.* Magnetotransport at Domain Walls in BiFeO3. *Phys. Rev. Lett.* **108**, 067203 (2012).

43. Gareeva, Z., Diéguez, O., Íñiguez, J. & Zvezdin, A. K. Complex domain walls in BiFeO3. *Phys. Rev. B* **91**, 060404 (2015).

44. Ghosal, A. *et al.* Low-energy domain wall racetracks with multiferroic topologies. *ArXiv* (2025) doi:arXiv:2507.12633.

45. Lebeugle, D. *et al.* Electric-field-induced spin flop in BiFeO3 single crystals at room temperature. *Phys. Rev. Lett.* **100**, 1–4 (2008).

46. Kibble, T. W. B. Topology of cosmic domains and strings. *J. Phys. A. Math. Gen.* **9**, 1387 (1976).

47. Zurek, W. H. Cosmological experiments in superfluid helium? *Nature* **317**, 505–508 (1985).

48. Jani, H. *et al.* Antiferromagnetic half-skyrmions and bimerons at room temperature. *Nature* **590**, 74–79 (2021).

49. Chmiel, F. P. *et al.* Observation of magnetic vortex pairs at room temperature in a planar α-Fe2O3/Co heterostructure. *Nat. Mater.* **17**, 581–585 (2018).

50. Urru, A. *et al.* G-type antiferromagnetic BiFeO3 is a multiferroic g-wave altermagnet. *Phys. Rev. B* **112**, 104411 (2025).

51. Šmejkal, L. Altermagnetic multiferroics and altermagnetoelectric effect. *ArXiv* 1–6 (2024).

52. Duan, X. *et al.* Antiferroelectric Altermagnets: Antiferroelectricity Alters Magnets. *Phys. Rev. Lett.* **134**, 106801 (2025).

53. Husain, S. *et al.* Anisotropic magnon transport in an antiferromagnetic trilayer heterostructure: is BiFeO3 an altermagnet? (2026).

54. Šmejkal, L., Sinova, J. & Jungwirth, T. Beyond Conventional Ferromagnetism and Antiferromagnetism: A Phase with Nonrelativistic Spin and Crystal Rotation Symmetry. *Phys. Rev. X* **12**, 031042 (2022).

55. Gu, M. *et al.* Ferroelectric Switchable Altermagnetism. *Phys. Rev. Lett.* **134**, 106802 (2025).

56. Hariki, A. *et al.* X-Ray Magnetic Circular Dichroism in Altermagnetic α -MnTe. *Phys. Rev. Lett.* **132**, 176701 (2024).

57. Amin, O. J. *et al.* Nanoscale imaging and control of altermagnetism in MnTe. *Nature* **636**, 348–353 (2024).

58. Galindez-Ruales, E. *et al.* Revealing the Altermagnetism in Hematite via XMCD Imaging and Anomalous Hall Electrical Transport. *Adv. Mater.* **37**, e05019 (2025).

59. Šmejkal, L. *et al.* Crystal time-reversal symmetry breaking and spontaneous Hall effect in collinear antiferromagnets. *Sci. Adv.* **6**, eaaz8809 (2020).

## Methods

*Molecular-Beam Epitaxy:* All films (thicknesses 2-50 nm) were grown by reactive MBE in a Veeco GEN10 system using a mixture of 80% ozone (distilled) and 20% oxygen. $SrRuO_3$ bottom electrodes were grown using an elemental source of Sr and an e-beam Ru source at fluxes of $1\times10^{13}$ atoms/$cm^2$s and $3\times10^{13}$ atoms/$cm^2$s respectively. Elemental sources of Bi and Fe were used at fluxes of $1.5\times10^{14}$ and $2\times10^{13}$ atoms/$cm^2$s, respectively, corresponding to a relative flux ratio of 8:1. To prevent Ru desorption from the surface, an initial $Fe_2O_3$ layer was deposited before co-deposition of both Bi and Fe. All films were grown at a substrate temperature of 675 °C as measured by an optical pyrometer operating at a wavelength of 980 nm and a chamber background pressure of $5\times10^{-6}$ Torr.

*Scanning Transmission electron microscopy:* The cross-section samples were prepared using a FEI Helios 660 focused ion beam (FIB) with a gallium (Ga) ion beam source. After sample preparation, high-angle annular dark-field scanning transmission electron microscopy (HAADF-STEM) imaging were performed using an FEI Titan Themis G3 (scanning) transmission electron microscope equipped with double Cs-correctors operated at 300 kV. The convergence semi-angle is 25 mrad and the collection angle is 48-200 mrad.

*Multislice Electron Ptychography:* The four-dimensional STEM (4D-STEM) dataset was collected using a high dynamic-range electron microscope pixel array detector (EMPAD) with 128 x 128 pixel[60]. The scan step size is 0.349 Å and the acquisition time per diffraction pattern is 1 ms with an additional 0.86 ms readout time. A probe defocus of 10 nm was used to increase the real-space probe overlap area for data redundancy. The mixed-state, multislice electron ptychography method is based on a customized PtychoShelves package[61], fold-slice[62–65], to reconstruct phase information. Bayesian optimization[66] was first used to determine the defocus and sample thickness for generating the initial probe. Then the "Ptychographic Experiment and Analysis Robot" (PEAR)[67], a framework that leverages large language models (LLMs), was used to assist with the processing of the data.

*Piezoresponse Force Microscopy:* The ferroelectric domains in BFO were imaged by piezoresponse force microscopy (PFM) using Asylum Jupiter XR and MFP-3D Origin atomic force microscopes. Out-of-plane components of the polarization were imaged in DART mode and in-plane components were imaged in lateral mode. Electrical poling of the samples and subsequent verification of ferroelectric behavior were performed using custom hardware controlled via the PIEC Python package[68].

*Quadrature phase differential interferometer-based PFM:* We used a commercially available Vero AFM system (from Asylum Research, an Oxford Instruments company) equipped with a quadrature phase differential interferometer (QPDI). We employed ADAMA commercially available single-crystal diamond probes (model AD-2.8-AS). The "null" spot of the cantilever, which mitigates electrostatic and cantilever dynamic artifacts, is found by moving the laser spot until the in-contact resonance frequency peak disappears, as explained in detail[69].

*Scanning Nitrogen Vacancy Magnetometry:* NV magnetometry measurements were performed using a Qnami Quantum Microscope–ProteusQ™. Parabolically tapered Quantilever™ MX+ diamond tips were employed for their high photon collection efficiency and superior signal-to-noise ratio, making them ideally suited for detecting the extremely weak stray magnetic fields characteristic of BFO. Each tip hosts a single negatively charged nitrogen–vacancy ($NV^-$) center, consisting of a substitutional nitrogen atom adjacent to a carbon vacancy. The electronic ground state of the NV center is a spin triplet ($m_s = 0, \pm 1$), which forms the basis of the quantum sensor. In the experimental configuration, an external ring magnet is applied to lift the degeneracy of the $m_s = \pm 1$ sublevels. A microwave (MW) source is then swept in frequency to induce spin transitions ($m_s = 0 \leftrightarrow \pm 1$), which are detected optically via changes in the photoluminescence (PL) intensity, an effect known as optically detected magnetic resonance (ODMR). As the NV tip scans across the sample surface, local variations in the stray magnetic field (B) projected along the NV axis shift the ODMR resonance frequencies, thereby providing a spatial map of the magnetic field distribution.

Two imaging modes were employed in this study: the dual iso-B mode and the full-B mode. In the dual iso-B mode, the PL signal is recorded at two fixed MW frequencies chosen near the full width at half maximum (FWHM) of the ODMR line shape, and their difference [PL($\nu_2$) – PL($\nu_1$)] yields a high-contrast map of the local magnetic texture. In the full-B mode, the complete ODMR spectrum is acquired and fitted at each pixel, allowing a quantitative reconstruction of the local magnetic field magnitude with nanoscale spatial resolution.

The typical magnetic field sensitivity achievable under continuous-wave ODMR conditions with Quantilever™ MX+ diamond tips was on the order of 1.5–2.5 μT $Hz^{-1/2}$. This limit arises from the interplay of shot-noise-limited photon detection, PL contrast, and laser- and microwave-induced broadening, and is consistent with the theoretical limit for this system[70]. To overcome this limitation in the thinnest, 1.6 nm BFO sample, we also implemented pulsed ODMR measurements, wherein a π-pulse is used to drive the spin transition, while the MW field remains off during the spin readout window. This protocol eliminates laser-induced broadening and contrast degradation, allowing us to clearly resolve the hyperfine structure associated with the $^{14}N$ nuclear spin. For quantitative imaging, we deliberately selected the central hyperfine resonance to ensure uniform signal response and optimal field sensitivity across the scan area.

*Laboratory-based X-ray diffraction:* To verify epitaxial growth, reciprocal space map (RSM) scans were performed on a Panalytical diffractometer with a Cu K-alpha source. RSMs were taken around the (332) DSO diffraction peak along the $[001]_O$ axis to access the $(103)_{PC}$ BFO peaks.

*Macroscopic XMLD Spectroscopy*: Macroscopic X-ray Linear Dichroism (XLD) measurements were taken at the Advanced Light Source (ALS) at Lawrence Berkeley National Laboratory (LBNL) at beamline 4.0.2 at the Fe $L_2$-edge. Measurements were performed at both normal (90°) incidence and grazing (16°) incidence. We quantify the dichroic signal as $\text{XLD Intensity} = (I_p - I_s)/(I_p + I_s)$, where $I_p$ and $I_s$ are total electron yield (TEY) intensities for $p$- and $s$-polarized light.

*Spatially resolved XMLD and XMCD PEEM Imaging:* The X-ray PEEM measurements (and corresponding XMLD) were taken at the MAXPEEM end-station at the MAXIV synchrotron. The X-ray beam is incident normal to the sample surface, with the X-ray linear vector in the plane of the sample and the helicity vector out of plane. The X-Ray Magnetic Linear Dichroism is calculated using $\text{XLD Intensity} = \left(I(E_1) - I(E_2)\right)/\left(I(E_1) + I(E_2)\right)$, where $I$ refers to the pixel intensity measured at the corresponding energies. For our case, $E_1, E_2$ corresponds to the two peaks of the L3 (or L2) absorption peak (Supplementary Fig. 15). The XMCD, magnetic circular dichroism, was calculated via $XMCD = \left(I(\mu_+) - I(\mu_-)\right)/\left(I(\mu_+) + I(\mu_-)\right)$, where $I$ refers to the pixel intensity measured at the corresponding helicities $\mu_\pm$ at a fixed energy. All measurements were taken at room temperature.

XMLD vector maps were created from a set of XMLD images ranging from $\theta = -90°$ to $\theta = 90°$ relative to the horizontal axis in steps of $10°$. The angular dependance is fitted with $A\,sin^2(\theta + \phi) + B$, where $\phi$ (the phase offset) describes the local Néel-vector axis (Supplementary Fig. 16). We further plot the histogram of the binned pixel values from the fit in a polar plot to illustrate the contributions to the XMLD signal (Supplementary Discussion VI) as in Figure 3. We note that we cannot resolve 180° reversal; i.e., the polar plot only contains information in 180° hemisphere. Finally, we binarize the XMLD (two states), then split each state by the sign of XMCD at that pixel, resulting in 2 unique states that represent the absolute orientation of the Néel vector oriented along $[100]_{PC}$, with $[010]_{PC}$ remaining undetermined due to symmetry constraints (Supplementary Figs. 17-19). Data analysis was done using the *athina* package[71], as well as custom python functions. We note that crystal fields can also give rise to a XLD signal. This can be neglected here as the crystal structure is in approximation cubic, similar to previous studies on hematite, the parent compound of BFO[48,49].

*First-Principles Calculations:* Calculations were performed using DFT[72] with the projector augmented wave (PAW) method[73] as implemented in the Vienna *ab initio* simulation package (VASP 5.4.4)[74]. The structural relaxations are carried out using a Γ-centered k-point grid equivalent to an 8x8x8 sampling of the five-atom unit cell Brillouin zone. The energy cutoff for the plane wave basis is set to 600 eV. The PAW potential corresponds to $5d^{10}\ 6s^{2}\ 6p^{3}$ valence electron configuration for Bi**,** $3p^{6}\ 3d^{7}\ 4s^{1}$ for Fe, $4p^{6}\ 4d^{7}\ 5s^{1}$ for Ru as well as $4s^{2}\ 4p^{6}\ 4d^{0.001}\ 5s^{1.999}$ and $2s^{2}\ 2p^{4}$ for Sr and oxygen, respectively. We employ the PBE + U and PBEsol + U functional form of the generalized gradient approximation[75], with a commonly used values of $U_{eff}$ = 4 eV and 0.6 eV for the Fe and Ru 3d orbitals[76,77] respectively, according to Dudarev's approach[78].

For structural relaxation in BFO/SRO bilayer setup, we also use Fermi smearing with smearing width $\sigma = 0.15$ eV and introduce electrostatic corrections to reduce the leading errors due to non-zero dipole moment under periodic boundary conditions[79,80]. We do not impose any symmetry constraints during structural and electronic relaxations but fix the geometry of the three bottom-most SRO unit cell layers according to HAADF-STEM observations.

*Spin- and angle-resolved photoemission spectroscopy (spin-ARPES)*: All in-situ photoemission measurements were conducted by immediately transferring the samples through a UHV manifold ($P < 1 \times 10^{-9}$ Torr) to a measurement chamber immediately following film growth. ARPES measurements were taken with a Scienta Omicron DA30-L electron analyzer equipped with a Fermion Instruments BL1200s plasma discharge lamp using He-Iα photons at 21.2 eV. The base pressure in the ARPES system is better than $5 \times 10^{-11}$ Torr. The sample was held at 300 K during measurements. To avoid sample charging during ARPES measurements, the film was grounded using retractable contact pins built onto the sample manipulator. For spin integrated measurements, the energy resolution was nominally set at 20 meV. Spin resolved data were taken using a VLEED Ferrum spin detector (Scienta Omicron), with the energy resolution during SRARPES nominally set at 40 meV. The polarization was calculated as the normalized intensity asymmetry measured for the oppositely poled magnetization states of the detector (I+, I−) divided by the instrument-specific Sherman function of S = 0.25. Typically short (< 1 nm) electron escape depths at VUV ARPES kinetic energies ensure that our measurements reflect only the intrinsic properties of the (1.6 nm thick) BFO top layer. This is further confirmed by the absence of any observed $SrRuO_3$ band structure features or spectral weight within 1eV of $E_F$ (Extended Data Fig 8).

*XMCD spectra simulations:* To calculate Fe $L_{2,3}$-edge XMCD of BFO, we employ an $Fe^{3+}$ ionic model that includes a molecular field simulating the altermagnetic order with a small canted ferromagnetic moment. The ionic model can describe the XMCD line shape reasonably well in altermagnets with localized 3d electrons, as demonstrated in recent studies on α-MnTe and α-$Fe_2O_3$ [56,81]. We use the same implementation adopted in these studies. First, DFT calculations within the local density approximation (LDA) were performed using the WIEN2k package[82], followed by the construction of a tight-binding (TB) Hamiltonian spanning the Fe 3d bands[83,84]. The ionic-model Hamiltonian includes the on-site crystal-field terms of the TB Hamiltonian and further incorporates the electron-electron interaction within the Fe 3d shell, parameterized by Hund's coupling J = 1.0 eV. In the XMCD simulations, the core-valence (2p–3d) multiplet interaction is explicitly included. The spin-orbit coupling (SOC) within both the valence (3d) and core (2p) shells is also taken into account. The valence-valence and core-valence interaction parameters, as well as the SOC parameters, are set to the same values used for α-$Fe_2O_3$[81,85], a localized $Fe^{3+}$ insulating system closely related to BFO.

**Data Availability:** Data presented in the Main Text and Extended Data can be found in a public repository on Zenodo[86].

**Code Availability**: Previously published software used to analyze X-ray dichroism, transmission electron microscopy, and nitrogen-vacancy magnetometry data can be found in the following repositories, as well as the manuscript Zenodo public repository[86]: https://gitlab.com/evangelosgolias/athina,

https://github.com/yijiang1/fold_slice, https://github.com/FangLinF/CalAtom, https://github.com/paulstevensonlab/snvm-processing.

**Methods-only References:**

60. Tate, M. W. *et al.* High Dynamic Range Pixel Array Detector for Scanning Transmission Electron Microscopy. *Microsc. Microanal.* **22**, 237–249 (2016).

61. Wakonig, K. *et al.* PtychoShelves, a versatile high-level framework for high-performance analysis of ptychographic data. *J. Appl. Crystallogr.* **53**, 574–586 (2020).

62. Thibault, P. *et al.* High-Resolution Scanning X-ray Diffraction Microscopy. *Science (80-. ).* **321**, 379–382 (2008).

63. Suzuki, A., Niida, Y., Joti, Y., Thibault, P. & Guizar-Sicairos, M. Maximum-likelihood refinement for coherent diffractive imaging. *New J. Phys.* **14**, 063004 (2012).

64. Thibault, P. & Menzel, A. Reconstructing state mixtures from diffraction measurements. *Nature* **494**, 68–71 (2013).

65. Usov, I., Tsai, E. H. R., Guizar-Sicairos, M., Menzel, A. & Diaz, A. X-ray ptychography with extended depth of field. *Opt. Express, Vol. 24, Issue 25, pp. 29089-29108* **24**, 29089–29108 (2016).

66. Cao, M. C., Chen, Z., Jiang, Y. & Han, Y. Automatic parameter selection for electron ptychography via Bayesian optimization. *Sci. Rep.* **12**, 1–10 (2022).

67. Yin, X., Shi, C., Deng, J., Han, Y. & Jiang, Y. PEAR: A Knowledge-guided Autonomous Pipeline for Ptychography Enabled by Large Language Models. *Microsc. Microanal.* **30**, 402–403 (2024).

68. Fratian, G., Qualls, A., Phan, J., Pankaj, R. & Caretta, L. PIEC: A Python Library for Integrated Experimental Control [Software].

69. Collins, L., Liu, Y., Ovchinnikova, O. S. & Proksch, R. Quantitative Electromechanical Atomic Force Microscopy. *ACS Nano* **13**, 8055–8066 (2019).

70. Dréau, A. *et al.* Avoiding power broadening in optically detected magnetic resonance of single NV

defects for enhanced dc magnetic field sensitivity. *Phys. Rev. B* **84**, 195204 (2011).

71. Golias, E. athina. (2024).

72. Kohn, W. & Sham, L. J. Self-Consistent Equations Including Exchange and Correlation Effects. *Phys. Rev.* **140**, A1133 (1965).

73. Blöchl, P. E. Projector augmented-wave method. *Phys. Rev. B* **50**, 17953 (1994).

74. Kresse, G. & Furthmüller, J. Efficient iterative schemes for ab initio total-energy calculations using a plane-wave basis set. *Phys. Rev. B* **54**, 11169 (1996).

75. Perdew, J. P., Burke, K. & Ernzerhof, M. Generalized Gradient Approximation Made Simple. *Phys. Rev. Lett.* **77**, 3865 (1996).

76. Neaton, J. B., Ederer, C., Waghmare, U. V., Spaldin, N. A. & Rabe, K. M. First-principles study of spontaneous polarization in multiferroic BiFeO 3. *Phys. Rev. B - Condens. Matter Mater. Phys.* **71**, 014113 (2005).

77. Shenton, J. K., Bowler, D. R. & Cheah, W. L. Effects of the Hubbard U on density functional-based predictions of BiFeO3 properties. *J. Phys. Condens. Matter* **29**, 445501 (2017).

78. Dudarev, S. L., Botton, G. A., Savrasov, S. Y., Humphreys, C. J. & Sutton, A. P. Electron-energy-loss spectra and the structural stability of nickel oxide: An LSDA+U study. *Phys. Rev. B* **57**, 1505–1509 (1998).

79. Neugebauer, J. & Scheffler, M. Adsorbate-substrate and adsorbate-adsorbate interactions of Na and K adlayers on Al(111). *Phys. Rev. B* **46**, 16067 (1992).

80. Makov, G. & Payne, M. C. Periodic boundary conditions in ab initio calculations. *Phys. Rev. B* **51**, 4014 (1995).

81. Yamamoto, R. *et al.* Altermagnetic nanotextures revealed in bulk MnTe. *Phys. Rev. Appl.* **24**, 034037 (2025).

82. Blaha, P. *et al.* WIEN2k: An APW+lo program for calculating the properties of solids. *J. Chem. Phys.* **152**, (2020).

83. Kuneš, J. *et al.* Wien2wannier: From linearized augmented plane waves to maximally localized

Wannier functions. *Comput. Phys. Commun.* **181**, 1888–1895 (2010).

84. Mostofi, A. A. *et al.* An updated version of wannier90: A tool for obtaining maximally-localised Wannier functions. *Comput. Phys. Commun.* **185**, 2309–2310 (2014).

85. Li, J. *et al.* Single- and Multimagnon Dynamics in Antiferromagnetic alpha-Fe2O3 Thin Films. *Phys. Rev. X* **13**, 011012 (2023).

86. Fratian, G. *et al.* Altermagnetic Signatures and Topological Textures in Ultrathin Multiferroic BiFeO3: Manuscript Data. *Zenodo* (2026) doi:10.5281/ZENODO.22697465.

**Figure 1:**

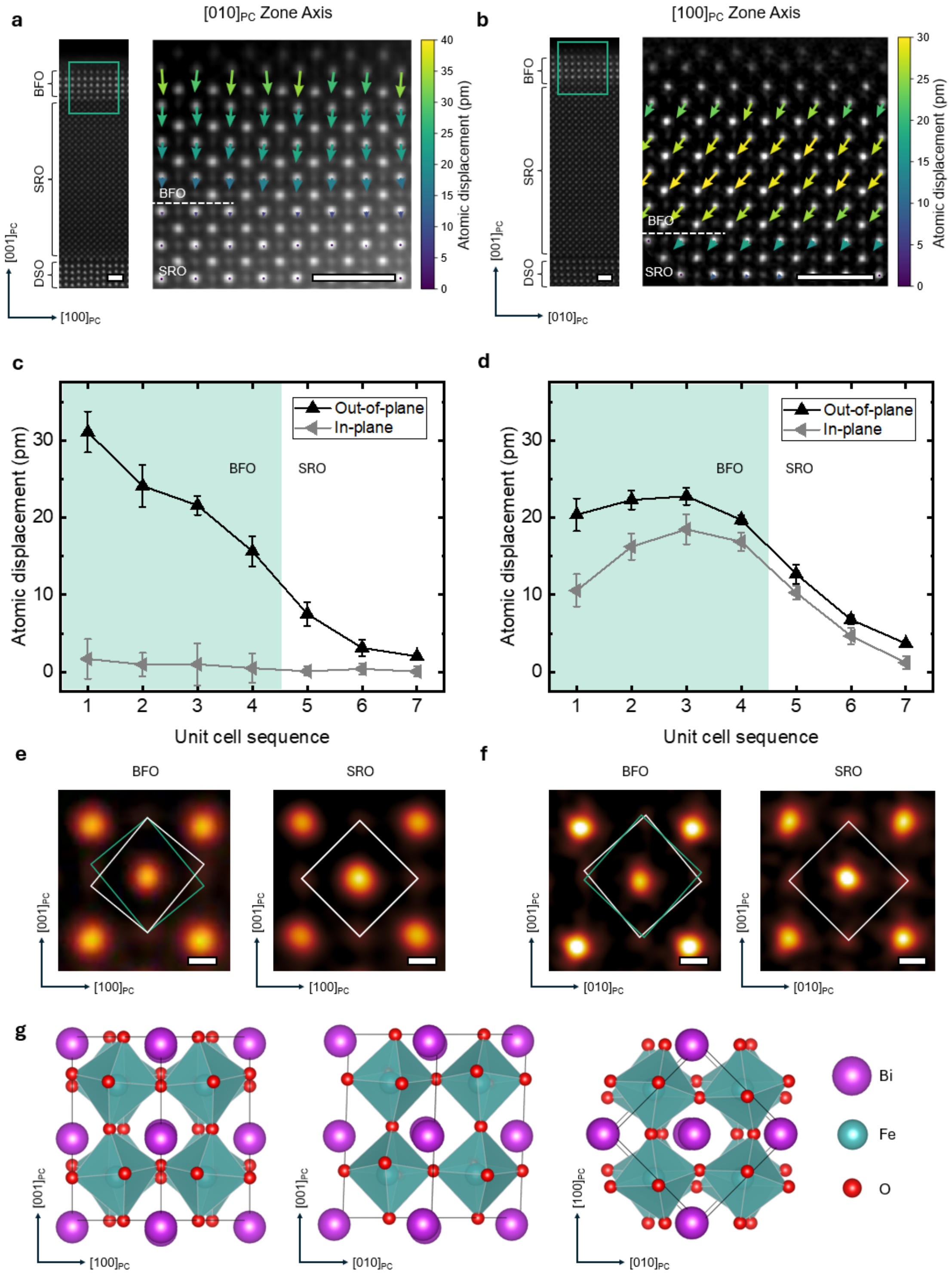
a
[010]PC Zone Axis
BFO
SRO
DSO
[001]PC
[100]PC
Atomic displacement (pm)
b
[100]PC Zone Axis
[010]PC
c
Out-of-plane
In-plane
Unit cell sequence
d
e
f
g
Bi
Fe
O

**Figure 2:**

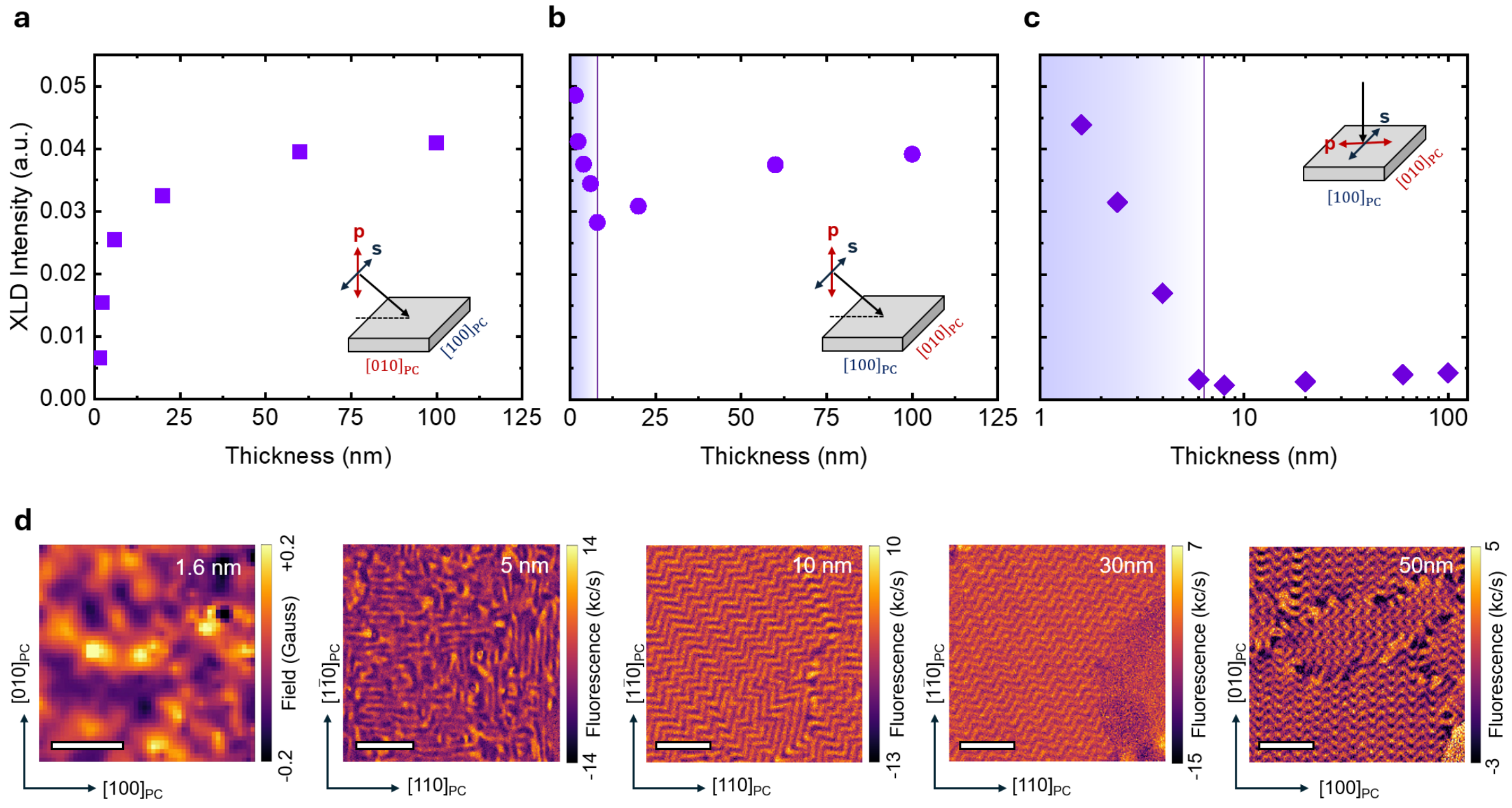

a
b
c
d
XLD Intensity (a.u.)
Thickness (nm)
Thickness (nm)
Thickness (nm)
p
s
[010]PC
[100]PC
[100]PC
[010]PC
1.6 nm
5 nm
10 nm
30nm
50nm
Field (Gauss)
Fluorescence (kc/s)
[100]PC
[110]PC
[010]PC
[1$\bar{1}$0]PC

**Figure 3:**

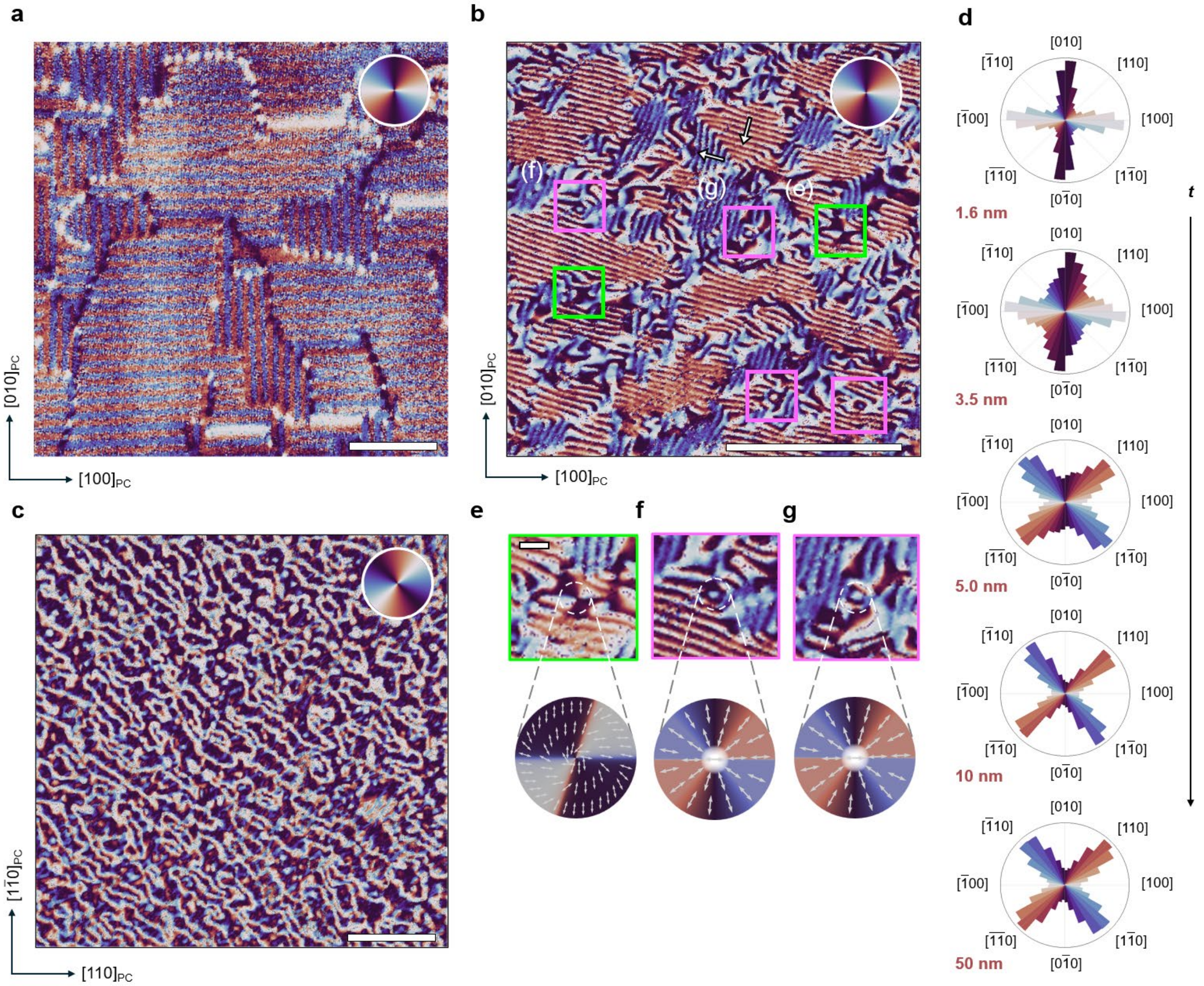
a
b
c
d
e
f
g
[010]PC
[100]PC
[1$\bar{1}$0]PC
[110]PC
(f)
(g)
(e)
[010]
[$\bar{1}$10]
[110]
[$\bar{1}$00]
[100]
[$\bar{1}\bar{1}$0]
[1$\bar{1}$0]
[0$\bar{1}$0]
1.6 nm
3.5 nm
5.0 nm
10 nm
50 nm
t

**Figure 4:**

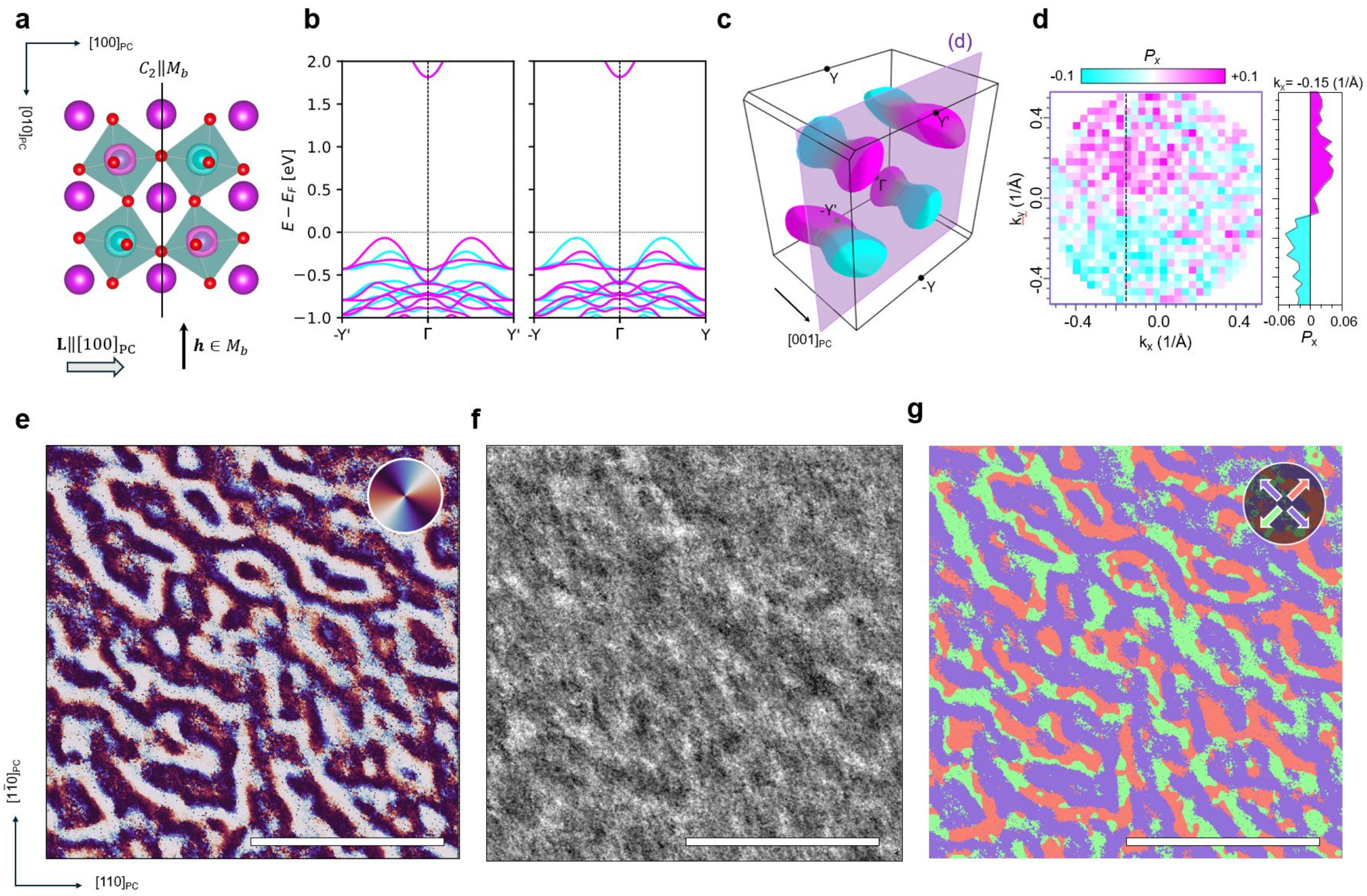
a
b
c
d
e
f
g
[100]PC
[010]PC
$C_2 \| M_b$
$\mathbf{L} \| [100]_{PC}$
$\boldsymbol{h} \in M_b$
$E - E_F$ [eV]
-Y'
Γ
Y'
-Y
Y
(d)
[001]PC
$P_x$
-0.1
+0.1
$k_x$= -0.15 (1/Å)
$k_x$ (1/Å)
$k_y$ (1/Å)
[1$\bar{1}$0]PC
[110]PC